# Essential discreteness in generalized thermostatistics with non-logarithmic entropy

SUMIYOSHI ABE [1, 2, 3 (a)]

[1] *Department of Physical Engineering, Mie University, Mie 514-8507, Japan*

[2] *Institut Supérieur des Matériaux et Mécaniques Avancés, 44 F. A. Bartholdi, 72000 Le Mans, France*

[3] *Inspire Institute Inc., McLean, Virginia 22101, USA*

**Abstract**   It is shown by simple and straightforward considerations that discreteness of basic physical variables is, at least, essential for generalized statistical mechanics with non-logarithmic entropy to be thermodynamically applicable to classical systems. As a result, continuous Hamiltonian systems with long-range interactions and the so-called *q*-Gaussian momentum distributions are seen to be outside the scope of nonextensive statistical mechanics.

______________________

(a) E-mail: suabe@sf6.so-net.ne.jp

In the last two decades, a lot of efforts have been devoted to generalized thermostatistics in order to examine if it can describe stationary states of a class of Hamiltonian systems such as those with long-range interactions, which reside outside the realm of Boltzmann-Gibbs statistical mechanics. In this stream, nonextensive statistical mechanics [1] has been playing a central role, in particular.

Nonextensive statistical mechanics is a theory based on the $q$-entropy [1]:

$$S_q[p] = \frac{k_B}{1-q}\left[\sum_{i=1}^{W}(p_i)^q - 1\right]. \tag{1}$$

Here, $k_B$ is the Boltzmann constant, $W$ the total number of microscopically accessible states, $\{p_i\}_{i=1,2,...,W}$ a probability distribution. The entropic index, $q$, should be positive in order for $S_q$ not to be infinitely sensitive to infinitesimal probabilities. In marked contrast to the Boltzmann-Gibbs entropy

$$S[p] = -k_B \sum_{i=1}^{W} p_i \ln p_i, \tag{2}$$

it is of the non-logarithmic type for $q$ different from unity (the $q$-entropy converges to the Boltzmann-Gibbs entropy in the limit $q \to 1$). The associated statistical-mechanical distribution in canonical theory is customarily obtained by maximization of the $q$-entropy under the constraints on the normalization condition, $\sum_{i=1}^{W} p_i = 1$, and the internal energy, $U \equiv \langle H \rangle = \sum_{i=1}^{W} \varepsilon_i p_i$. The resulting stationary distribution reads [2]

$$\tilde{p}_i = \frac{1}{Z_q(\beta)}\left[1 - \frac{q-1}{qc_q}\beta(\varepsilon_i - \tilde{U})\right]_+^{1/(q-1)}, \tag{3}$$

where

$$Z_q(\beta) = \sum_{i=1}^{W} \left[ 1 - \frac{q-1}{qc_q} \beta (\varepsilon_i - \tilde{U}) \right]_+^{1/(q-1)}. \quad (4)$$

In these equations, $\beta$ is the Lagrange multiplier associated with the constraint on the internal energy, $[a]_+ \equiv \max\{a, 0\}$, $c_q \equiv \sum_{i=1}^{W} (\tilde{p}_i)^q$, and $\tilde{U}$ is calculated with respect to the stationary distribution in Eq. (3) in the self-referential manner. Here, we are purposely employing the discrete notation. For example, $\varepsilon_i$ stands for the *i*th value of the system Hamiltonian, *H*. Eq. (3) is referred to as the *q*-exponential distribution. It decays as a power law if $0 < q < 1$ (i.e., the asymptotic scale invariance), and tends to the Boltzmann-Gibbs exponential distribution in the limit $q \to 1$.

An information entropy in the maximum entropy principle is not necessarily the physical entropy. This issue has repeatedly been discussed in the literature. Consequently, the following consensus has been reached at a certain level: an information entropy can be regarded as the physical entropy *if and only if the whole framework and concepts are consistent with thermodynamics*. It is known [3] that the quantity in Eq. (2) is physical.

In this paper, we show that, in order for the *q*-entropy to be physical, systems to be treated should be *at least* discrete, such as spins on a lattice. The point is concerned with an interrelationship between additivity of the physical entropy, classical statistical mechanics and quantum theory. As a result, Hamiltonian systems with continuous canonical variables and the so-called *q*-Gaussian [1] momentum distribution are seen to be outside the scope of nonextensive statistical mechanics.

We start our discussion with the second law of thermodynamics. This law states that the change of the physical entropy, S, between the initial and final equilibrium states along a thermodynamic process satisfies

$$S^{final} - S^{initial} \geq \int_{initial}^{final} \frac{d'Q}{T}, \tag{5}$$

where Q and T are the quantity of heat and temperature, respectively. In nonextensive statistical mechanics, what is relevant is nonequilibrium stationary states, and not strictly equilibrium states. So, rigorously speaking, the thermodynamic concepts are somewhat vague, there. It is, however, a common understanding in this field that the laws of thermodynamics can naturally be generalized to nonequilibrium stationary states, since they usually survive for a long period of time, much longer than typical time scales of underlying microscopic dynamics, in general [4].

As known well, one of implications of Eq. (5) is that what is classically observable is not the absolute value of the entropy itself but its change. Although the third law of thermodynamics sets the absolute value of the entropy, that law is concerned with the low temperature limit, in which quantum effects play dominant roles.

Now, an information entropy is originally defined for finite discrete systems. In fact, the sets of axioms for the quantity of the form in Eq. (2), i.e., the Shannon entropy [5,6], and the $q$-entropy [7,8] are constructed based on this premise, and then the uniqueness theorems of these entropies are established [5-8]. When the maximum entropy principle is applied to a continuous system, a careful treatment is needed for the continuum limit of an information entropy. Let us consider a set of discrete points, $\{x_1, x_2, ..., x_n\}$, which divides a fixed interval $[a, b]$ as $a = x_1 < x_2 < \cdots < x_n = b$, and increase the number of points. Then, introduce a measure, $m$, as follows:

$$\frac{1}{m(x_i)} = \lim_{n \to \infty} n\,(x_{i+1} - x_i). \tag{6}$$

In the limit $n \to \infty$, the summation operation, $\sum_{i=1}^{n} [n\,m(x_i)]^{-1}$, becomes the integration operation, $\int_a^b dx$. And the probability distribution, $p_i$, and the probability density, $\rho(x_i)$, are related to each other as follows:

$$p_i = \rho(x_i)(x_{i+1} - x_i)$$

$$\to \rho(x_i) \frac{1}{n\,m(x_i)}. \tag{7}$$

Accordingly, $S$ in Eq. (2) tends to

$$S[p] \to -k_B \int_a^b dx\,\rho(x) \ln \frac{\rho(x)}{n\,m(x)} \qquad (n \to \infty). \tag{8}$$

Then, recalling the crucial point that what is classically observable is the entropy change, we can define

$$S[\rho] = -k_B \int_a^b dx\,\rho(x) \ln \frac{\rho(x)}{m(x)} \tag{9}$$

as the entropy for a continuous variable by discarding the additive logarithmic divergence, $\lim_{n \to \infty} k_B \ln n$, which is always cancelled in the entropy change. In the special case when the measure is zero, Eq. (7) apparently fails to define the probability density. However, such a difficulty is overcome in statistical mechanics with the help of the concept of coarse graining.

It may also be worth mentioning [9] that the measure, $m(x)$, is essential for Eq. (9) to be invariant under transformations of the variable.

The naive scheme mentioned above, however, breaks down in the case of the $q$-entropy in Eq. (1). The limit $n \to \infty$ of

$$S_q = \frac{k_B}{1-q}\left\{\int_a^b dx\, \rho(x)\left[\frac{\rho(x)}{n\, m_q(x)}\right]^{q-1} - 1\right\}, \tag{10}$$

with $m_q(x)$ being a measure in a relevant nonextensive system, does not yield any meaningful results. If $q > 1$, then $S_q$ tends to $k_B/(q-1)$, which is independent of the distribution. On the other hand, difference of $S_q$ diverges as well as $S_q$ itself if $0 < q < 1$, since the divergence is not additive but multiplicative. This fact brings a serious difficulty to defining the $q$-entropy for continuous variables.

In what follows, we examine if nonextensive statistical mechanics is thermodynamically applicable for Hamiltonian systems with continuous canonical variables.

Consider $\Gamma$ space, which is a $6N$-dimensional phase space $(\mathbf{p}_1, \mathbf{p}_2, ..., \mathbf{p}_N, \mathbf{q}_1, \mathbf{q}_2, ..., \mathbf{q}_N)$, where $N$ is the number of particles contained in the system, and $\mathbf{p}_i$ and $\mathbf{q}_i$ are the canonical momentum and coordinate of the $i$th particle, respectively. What is relevant is the constant-energy hypersurface with tiny thickness, $\Sigma$, embedded in $\Gamma$. To develop statistical mechanics, it is necessary to coarse grain each phase plane with the scale, $h$, which is the Planck constant. Let $\Omega$ be the phase-space volume occupied by the system. Then, the standard classical entropy is given by

$$S = k_B \ln \frac{\Omega}{h^{3N}}. \tag{11}$$

Thus, the Planck constant is indispensable though it is *arbitrarily small* in classical theory. However, the point of importance is that, due to the logarithmic nature, it never appears in the entropy change, the specific heat, $C_V = T(\partial S / \partial T)_{V,N}$, or the equation of state, $P = T(\partial S / \partial V)_{E,N}$, where $E$ and $V$ are the energy and volume of the system, respectively. The one and only exception is the chemical potential, $\mu = -T(\partial S / \partial N)_{E,V}$. Classically, however, the Planck constant is absorbed into the definition of the thermal wavelength, $\lambda_T$, which is typically $\lambda_T \propto h / \sqrt{T}$ and is a quantity to be experimentally determined. The classical limit corresponds to the short-wavelength limit, and consequently, classical theory is self-consistently valid in this way.

The situation drastically alters in the case of the *q*-entropy, because of its form

$$S_q = \frac{k_B}{1-q}\left[\left(\frac{\Omega_q}{h^{3N}}\right)^{1-q} - 1\right]. \qquad (12)$$

Here, $\Omega_q$ stands for the contracted phase-space volume. It is supposed to be significantly smaller than $\Omega$ in Eq. (11). The reason is that nonextensive systems in nonequilibrium stationary states involve strong correlations between elements and ergodicity in $\Sigma$ may be broken, in general. Accordingly, $\Sigma$ is not visited by the system, uniformly. (Discussions relevant to this consideration can be found, for example, in Refs. [10,11]). With the form in Eq. (12), it follows that all of the entropy change, the specific heat, and the equation of state explicitly depend on the Planck constant. This result for classical systems is not acceptable from the physical viewpoint.

A similar difficulty is present also in canonical theory. Consider the continuous counterparts of Eqs. (3) and (4):

$$\tilde{\rho}(\mathbf{p}_1, \mathbf{p}_2, ..., \mathbf{p}_N, \mathbf{q}_1, \mathbf{q}_2, ..., \mathbf{q}_N) = \frac{1}{Z_q(\beta)} \left[ 1 - \frac{q-1}{q\, c_q} \beta (H - \tilde{U}) \right]_+^{1/(q-1)}, \quad (13)$$

$$Z_q(\beta) = \int d\Gamma \left[ 1 - \frac{q-1}{q\, c_q} \beta (H - \tilde{U}) \right]_+^{1/(q-1)}, \quad (14)$$

where $H = H(\mathbf{p}_1, \mathbf{p}_2, ..., \mathbf{p}_N, \mathbf{q}_1, \mathbf{q}_2, ..., \mathbf{q}_N)$ is the system Hamiltonian and

$$d\Gamma \equiv \frac{d^3\mathbf{p}_1 d^3\mathbf{p}_2 \cdots d^3\mathbf{p}_N d^3\mathbf{q}_1 d^3\mathbf{q}_2 \cdots d^3\mathbf{q}_N}{h^{3N} g(N)} \quad (15)$$

with $g(N)$ being the Gibbs factor that may make the $q$-entropy extensive. The allowed range of $q$ is limited by the finiteness of the internal energy, $\tilde{U} = \int d\Gamma\, H\, \tilde{\rho}$. The $q$-entropy and $q$-free energy are given by

$$\tilde{S}_q = \frac{k_B}{1-q} \left\{ [Z_q(\beta)]^{1-q} - 1 \right\}, \quad (16)$$

$$F_q = \tilde{U} - \frac{1}{k_B \beta} \tilde{S}_q, \quad (17)$$

respectively. Again, due to the non-logarithmic nature of Eq. (16), all of the entropy change, the free-energy difference, the specific heat, and the equation of state explicitly contain the Planck constant [12], leading to physically unacceptable results in classical theory.

In the above discussion, $\beta^{-1}$ is regarded as $k_B T$ with $T$ being temperature. Actually, it is a nontrivial problem to identify what temperature is in nonextensive statistical mechanics [13].

Such a freedom, however, does not improve the situation, since the difficulties also come from the dimensional reason.

To summarize, we have shown that, for Hamiltonian systems with continuous canonical variables, nonextensive statistical mechanics is not thermodynamically applicable and the $q$-entropy fails to be physical, since macroscopic classical theory explicitly depends on the Planck constant. Therefore, the $q$-Gaussian distribution [1] [i.e., the quadratic dependence of the Hamiltonian on the canonical momenta in Eq. (13)] is irrelevant to nonextensive statistical mechanics, for example. The same result holds also for other non-logarithmic entropies such as the quantum-group entropy [14], the $\kappa$-entropy [15], and the homogeneous entropy [16]. Thus, we conclude that Hamiltonian systems to be treated by generalized thermostatistics with non-logarithmic entropy have to be, at least, discrete. This may physically be interpreted as an effect of the uncertainty principle, which strongly constrains the scale invariance in phase space.

\* \* \*

The author would like to thank A. K. Rajagopal for drawing his attention to Ref. [9]. This work was supported in part by a Grant-in-Aid for Scientific Research from the Japan Society for the Promotion of Science.

REFERENCES

[1] TSALLIS C., *Introduction to Nonextensive Statistical Mechanics: Approaching a Complex World* (Springer, New York) 2009.

[2] ABE S., *Phys. Rev. E*, **79**, (2009) 041116.

[3]  GRANDY JR. W. T., *Entropy and the Time Evolution of Macroscopic Systems* (Oxford University Press, Oxford) 2008.

[4]  LATORA V., RAPISARDA A. and TSALLIS C., *Phys. Rev. E*, **64** (2001) 056134.

[5]  SHANNON C. E. and WEAVER W., *The Mathematical Theory of Communication* (University of Illinois Press, Urbana) 1949.

[6]  KHINCHIN A. I., *Mathematical Foundations of Information Theory* (Dover, New York) 1957.

[7]  DOS SANTOS R. J. V., *J. Math. Phys.*, **38** (1997) 4104.

[8]  ABE S., *Phys. Lett. A*, **271** (2000) 74.

[9]  JAYNES E. T., Brandeis Lectures, reprinted in ROSENKRANTZ R. D. (Editor), *E. T. Jaynes: Papers on Probability, Statistics and Statistical Physics* (Kluwer, Dordrecht) 1989.

[10] GARCÍA-MORALES V. and PELLICER J., *Physica A*, **361** (2006) 161.

[11] OLEMSKOI A. I., KHARCHENKO V. O. and BORISYUK V. N., *Physica A*, **387** (2008) 1895.

[12] ABE S., *Phys. Lett. A*, **263** (1999) 424; erratum **267** (2000) 456.

[13] ABE S., *Physica A*, **368** (2006) 430.

[14] ABE S., *Phys. Lett. A*, **224** (1997) 326; *Phys. Lett. A*, **244** (1998) 229.

[15] KANIADAKIS G., *Phys. Rev. E*, **66** (2002) 056125; *Phys. Rev. E*, **72** (2005) 036108.

[16] LUTSKO J. F., BOON J. P. and GROSFILS P., *Europhys. Lett.*, **86** (2009) 40005.